\documentclass[12pt]{article}
\usepackage{amsmath,amssymb, bm, epsfig}
\newcommand{\lqqd}{\hfill $\square$\\}
\newcommand{\Z}{{\mathbb{Z}}}

\newtheorem{definition}{Definition}
\newtheorem{corollary}{Corollary}
\newtheorem{theorem}{Theorem}
\newtheorem{lemma}{Lemma}

\newtheorem{example}{Example}

\begin{document}
 \title{On the control of abelian group codes with information group of prime order }
\author{Jorge P Arpasi \thanks{arpasi@gmail.com}}
\maketitle
\begin{abstract}
Finite State Machine (FSM) model is widely used in the construction of binary convolutional codes.
If $\Z_2=\{0,1\}$ is the binary mod-2 addition group and $\Z_2^n$ is the $n$-times direct product of $\Z_2$,
then  a binary convolutional encoder, with rate
$\frac{k}{n}< 1$ and memory $m$, is a FSM with $\Z_2^k$ as inputs group, $\Z_2^n$ as outputs group 
and   $\Z_2^m$ as states group. The next state mapping $\nu:\Z_2^k\oplus \Z_2^m\rightarrow \Z_2^m$ is a surjective
group homomorphism. The encoding mapping $\omega:\Z_2^k\oplus \Z_2^m\rightarrow \Z_2^n$ is a homomorphism adequately
restricted by the trellis graph produced by $\nu$. The binary convolutional code is the
 family of bi-infinite sequences produced by the binary 
convolutional encoder. Thus, a convolutional code can be considered as a dynamical system 
and it is known that well behaved 
dynamical systems must be necessarily  controllable.

The generalization of binary convolutional encoders over arbitrary finite groups is made by  using the extension
of groups, instead of direct product. 
In this way, given finite groups $U,S$ and $Y$, a wide-sense homomorphic encoder (WSHE) is a FSM 
with $U$ as inputs group, $S$ as states group, and $Y$ as  outputs group.  By denoting $U\boxtimes S$ as the extension
of $U$ by $S$, the next state homomorphism  $\nu:U\boxtimes S\rightarrow S$ needs to be surjective 
and the encoding homomorphism  $\omega:U\boxtimes S\rightarrow Y$ has restrictions given by the trellis graph produced by
$\nu$. The code produced by a WSHE is known as group code.
In this work we will study the case when the extension $U\boxtimes S$ is abelian with $U$ being $\Z_p$, $p$ 
a positive prime number. We will show that this class of WSHEs  will produce controllable codes
only if the states group $S$ is isomorphic with $\Z_p^j$, for some positive integer $j$.
\end{abstract}

\textbf{keywords}
 Finite State Machine, Group Code, Dynamical System, Control.

\section{Introduction}

Group codes are a subclass of Error Correcting Codes (ECC), which can detect and correct transmission errors originated from noisy 
communication channels.
In communication engineering, noise is modeled as a random signal. The most known noise is the Gaussian noise, which is modeled as a 
random signal having a normal probabilistic distribution.
The channels suffering Gaussian noise are called \textit{additive white Gaussian noise} - AWGN channels \cite{moon,perez,mackay,haykin}.
The essence of an ECC is the addition of redundancy to the original message.
More redundant information means more protected information.
That fact reduces the transmission velocity of the channel. Then trade-off between velocity of transmission 
and protection of information must be done, and this depends on the channel class \cite{gallager,macwilliams}. 
Voice communication channels like  VOIP need real time transmissions and they prioritize velocity over 
some little errors on the human voice. 
On the other hand, bank transaction channels need strong protection on the transmitted data.

A special case of group codes are those generated by a particular class of Finite State Machines (FSM).
The encoder of a group code is a FSM $=(U,S,Y,\nu,\omega)$, where $U$
is the group of inputs, $S$ is the group of the FSM states and $Y$ is the group of outputs. 
The next state mapping
$\nu:U\boxtimes S\rightarrow S$ is a surjective homomorphism defined on the extension $U\boxtimes S$.
The encoding mapping $\omega:U\boxtimes S\rightarrow Y$
is a group homomorphism such that the mapp $(u,s)\mapsto (s,\omega(u,s),\nu(u,s))$ is one-to-one.
This FSM encoder is called Wide Sense Homomorphic Encoder (WSHE) \cite{loeliger, mittelholzer}. 
If $\Z$ is the set of integers, the  group code $\mathcal{C}$, generated by the WSHE,  is
a subgroup of $Y^{\Z}=\overbrace{\dots\times Y\times Y\times Y\times\dots}^{\Z\;times}$, where $Y$ is the group output of the WSHE.
This means that each element of a group code $\mathcal{C}\subset Y^{\Z}$ is a bi-infinite sequence
$\bm{y}=\{y_n\}_{n\in\Z}$, $y_n\in Y$. Hence $\mathcal{C}$ can be considered as a dynamical system in the sense
given by Willems in \cite{willems}. 
From the dynamical system point of view, a WSHE  
is the realization of the respective group code $\mathcal{C}$ that which we call FSM group code. A WSHE  is always linear and time invariant.
On the other hand, the behavior of a general group code is important because it has been shown,
for instance in \cite{mittelholzer,loeliger,trott,marcus}, that a good group code must be 
necessarily controllable. Here
the goodness of a group code is the coding gain, which 
is the measure in the difference between the signal to noise ratio (SNR) levels between the uncoded 
system and coded system required to reach the same bit error rate (BER).

In this paper we will deal with a subclass of WSHE where: \textbf{a)} the group extension $U\boxtimes S$ is
abelian and \textbf{b)} the input group $U$ is the cyclic group $\Z_p=\{1,2,\dots,p-1,p\}$ with $p$ prime.
We will show that an WSHE with these conditions will produce controllable group codes only
if $S$ is isomorphic with $\Z_p^j$, for some $j\geq 1$. For that, this work is organized as follows: \\
\underline{In the Section 2}  is defined the extension of a group $U$
by the group $S$, this extension is denoted as $U\boxtimes S$. It is shown that this group extension is a generalization of 
the known direct product and semi-direct product of groups.  Then is defined the WSHE of a group code
and it is exhibited practical techniques to generate b-infinite sequences of codewords generated by the WSHE.\\
\underline{In the Section 3} the group code $\mathcal{C}$ is presented as a set of trajectories
of a  dynamical system in the sense of \cite{willems}. Hence a group code generated by a WSHE is interpreted
as a dynamical system. The definition of controllable group codes is given. To have a more practical criterion
on deciding if a group code is controllable,  a graphical description, called trellis, of WSHE and its group codes
is given. Since the trellis elements are paths, it is shown that a group code will be controllable only if
any two states are connected by a finite path of its trellis. \\
\underline{In the Section 4} we present our original contributions. Thorough a sequence of Lemmas and Theorems
we will show, among other results, the following;
\begin{itemize}
 \item If $\Z_p\boxtimes S$ is abelian then $S$ must be abelian.
\item If $\Z_p\boxtimes \Z_m$ is abelian then either $\Z_p\boxtimes \Z_m$ is isomorphic with the direct product
 $\Z_p\boxtimes S$ or $\Z_p\boxtimes \Z_m$ is isomorphic with the cyclic group $\Z_{pm}$.
\item A WSHE defined over the abelian extension $\Z_p\boxtimes S$, with $S$ cyclic, will produce non-controllable 
group codes.
\item If a WSHE defined over the abelian extension $\Z_p\boxtimes S$ produce controllable 
controllable group codes then $S$ must have be isomorphic with 
$\Z_p^j=\overbrace{\Z_p\oplus\Z_p\oplus\dots\oplus\Z_p}^{j-times}$, 
for some natural number $j\geq 1$.
\end{itemize}

\section{Group extensions and Group Codes}
\begin{definition}\label{def:extension}
Given a group $G$ with a normal subgroup $N$ consider the quotient group $\frac{G}{N}$. If there are
two groups $U$ and $S$ such that $U$ is isomorphic with $N$ and $S$ is isomorphic with $\frac{G}{N}$
then it is said that $G$ is an \textbf{extension} of $U$ by $S$ \cite{rotman}. \lqqd
\end{definition}
The extension ``$U$ by $S$" we will denote by the symbol $U\boxtimes S$, also we will use the standard notations $U\cong N$
meaning ``$U$ is isomorphic with $S$'' and $N \triangleleft G$ meaning ``$N$ normal subgroup of $G$'' .
 When $G$ is an extension $U\boxtimes S$, each element $g\in G$ can be ``factored'' as an unique 
ordered pair $(u,s)$, $u\in U$ and $s\in S$.
The semi-direct product $U\rtimes S$ is a particular case of extension, but also it is known that the semi-direct
product  is a generalization of the direct product $U\times S$.  A canonical definition of extension of groups 
is given in  \cite{rotman,hall},
specially in \cite{hall} we find a  ``practical" way to decompose a given group $G$, with normal subgroup $N$, in
an extension $U\boxtimes S$.
That decomposition depends on the choice of  isomorphisms $\upsilon:N\rightarrow U$, $\psi:S\rightarrow \frac{G}{N}$
and a lifting $l:\frac{G}{N}\rightarrow G$ such that $l(N)=e$, the neutral element of $G$. 
Then, defining $\phi:S\rightarrow Aut(U)$ by,
\begin{equation}\label{eq:phi}
\phi(s)(u)=\upsilon[l(\psi(s))\cdot\upsilon^{-1}(u)\cdot(l(\psi(s)))^{-1}],
\end{equation}
and $\xi:S\times S \rightarrow U$
\begin{equation}\label{eq:xi}
\xi(s_1,s_2)=l(\psi(s_1s_2))(l(\psi(s_1)))^{-1}(l(\psi(s_2)))^{-1},
\end{equation}
the decomposition $U\boxtimes S$ with the group operation
\begin{equation}\label{eq:operation}
 (u_1,s_1)*(u_2,s_2)=(u_1\cdot\phi(s_1)(u_2)\cdot\xi(s_1,s_2)\,,\,s_1s_2)
\end{equation}
is isomorphic with $G$, that is, $g=(u,s)$.

Notice that the resulting  pair of $(u_1,s_1)*(u_2,s_2)$, of the above operation (\ref{eq:operation}), 
is $(u^\prime,s_1s_2)$ for some $u^\prime\in U$, and $s_1s_2$ is the operation on $S$.
This property allow us to do not be concerned to obtain an explicit result when multiple factors are acting.
 For instance, in the proof of some Lemmas it will be enough to say that  $(u^\prime,s_1s_2\dots s_n)$, is the resulting pair
of the multiple product $(u_1,s_1)*(u_2,s_2)*(u_3,s_3)*\dots*(u_n,s_n)$, where $u^\prime$ is some element 
of $U$. Analogously, $(u,s)^n=(u^\prime, s^n)$ for some $u^\prime \in U$.

\begin{example}\label{ex:binary}
Consider the direct product group $\Z_2^3=\{(x_1,x_2,x_3)\;;\;x_i\in \Z_2\}$. This abelian group can be
decomposed as an extension  $\Z_2\boxtimes \Z_2^2$. 
\end{example}

By using the more convenient notation $00$ instead $(0,0)$, $010$ instead $(0,1,0)$, etc.,  we have that
the normal subgroup $N=\{000,100\}\triangleleft \Z_2^3$ is isomorphic with $\Z_2$. The 
quotient group $\frac{\Z_2^3}{N}$ = $\{\{000,100\},\{010,110\},\{001,101\},\{111,011\}\}$ is
isomorphic with $\Z_2^2$. Thus, in an expected way, we have shown that $\Z_2^3$ is an extension of
$\Z_2\boxtimes \Z_2^2$. 

\begin{theorem}\label{theo:phi_nonabel}
 If the mapping $\phi:S\rightarrow Aut(U)$ is not trivial then the extension $U\boxtimes S$ is non-abelian
\end{theorem}
\textbf{Proof.-} Since $\phi$ is not trivial, there are  $u\in U$ and $s\in S$ such that $\phi(s)(u)\neq u$. Now, consider the
pairs $(e,s),(u,e)\in U\boxtimes S$, where $e$ is the neutral element of the respective group. Then $(e,s)*(u,e)$ =
$(e.\phi(s)(u).\xi(s,e),s)$ = $(\phi(s)(u),s)$. On the other hand $(u,e)*(e,s)$ = $(u.\phi(e)(e).\xi(e,s),s)$ = $(u,s)$.
Therefore $(e,s)*(u,e)\neq (u,e)*(e,s)$. \lqqd

\subsection{Group codes generated by finite state machines}
Finite state machines (FSM) are a subject of Automata Theory. M. Arbib in \cite{arbib} describes a FSM as 
a quintuple $M=(I,S,O,\delta,\xi)$, where $I$ is the inputs alphabet, $S$ is the alphabet
of states of the machine, $O$ is the outputs alphabet, $\delta:I\times S\rightarrow S$ is the next state mapping,
and $\xi:I\times S\rightarrow O$ is the output mapping. The encoder of a group code is subclass of FSM which
is called \textit{wide-sense homomorphic encoder} (WSHE) \cite{loeliger,mittelholzer}.
\begin{definition}\label{def:encoder}
A wide-sense homomorphic encoder (WSHE) is a  machine $M=(U,S,Y,\nu,\omega)$, where
$U$, $S$, and $Y$ are finite groups,  $\nu:U\boxtimes S\rightarrow S$ and $\omega:U\boxtimes S \rightarrow Y$ are
group homomorphisms  defined on an extension $U\boxtimes S$ such that the mapping $\nu$ is surjective and
$\Psi:U\boxtimes S\rightarrow S\times Y \times S$ defined by
\begin{equation}\label{eq:Psi}
\Psi(u,s)=(s,\omega(u,s),\nu(u,s))
\end{equation}
is injective.\lqqd
\end{definition} 
The group $U$ is called the uncoded information group, $Y$ is called the encoded information group, and
the group $S$ is the states group of the WSHE.
The WSHE generates a group code $\mathcal{C}\subset Y^\Z$ as follows: \\
\underline{\textbf{For future indexes.-}} Given an initial state $s_0\in S$ and a sequence of uncoded 
inputs $\{u_i\}_{i\in \Z^{(+)}}$, where $u_i\in U$ and
$\Z^{(+)}=\{1,2,3,\dots\}\subset \Z$; there
 is a \textbf{unique} sequence $\{y_i\}_{i\in \Z^{(+)}}$, $y_i\in Y$, of encoded outputs, which is the response of
the WSHE, by the following recurrence relations:
\begin{equation}\label{eq:d_system+}
\begin{array}{cc|cc}
 \nu(u_1,s_0)&=s_1,&\omega(u_1,s_0)&=y_1,\\
\nu(u_2,s_1)&=s_2,&\omega(u_2,s_1)&=y_2,\\
\nu(u_3,s_2)&=s_3,&\omega(u_3,s_2)&=y_3,\\
\vdots&\vdots&\vdots&\vdots \\
\nu(u_n,s_{n-1})&=s_n.&\omega(u_n,s_{n-1})&=y_n\\
\vdots&\vdots&\vdots&\vdots \\
\end{array}
\end{equation}
It can observed that  $\{y_i\}_{i=1}^n$ depends on $\{s_i\}_{i=1}^n$, but the converse is not true, $\{s_i\}_{i=1}^n$ does not depend
on $\{y_i\}_{i=1}^n$.\\
\underline{\textbf{For past indexes.-}} The past states $s_{-1}$, $s_{-2},$ etc. are chosen considering that
$\nu$  is surjective. Beginning by the initial state $s_0$, we have that there must exist,
 at least one pair, that we conveniently call,  $(u_0,s_{-1})$  such that $s_0=\nu(u_0,s_{-1})$.
Analogously for this  $s_{-1}$ there must exist a pair $(u_{-1},s_{-2})$ such that $\nu(u_{-1},s_{-2})=s_{-1}$,
etc. Thus, for a given present/initial state $s_0$, there are sequences of 
past states $\{s_i\}_{i\in \Z^{(-)}}$, past outputs  $\{\{y_i\}_{i\in \Z^{(-)}}\}\cup \{y_0\}$, and past inputs 
$\{\{u_i\}_{i\in\Z^{(-)}}\}\cup \{u_0\}$, where $\Z^{(-)}=\{\dots,-3,-2,-1\}\subset \Z$,
such that:
\begin{equation}\label{eq:d_system-}
 \begin{array}{cc|cc}
 \nu(u_0,s_{-1})&=s_{0},&\omega(u_0,s_{-1})&=y_0,\\
\nu(u_{-1},s_{-2})&=s_{-1},&\omega(u_{-1},s_{-2})&=y_{-1},\\
\nu(u_{-2},s_{-3})&=s_{-2},&\omega(u_{-2},s_{-3})&=y_{-2},\\
\vdots&\vdots& \vdots&\vdots \\
\nu(u_{\{-n+1\}},s_{-n})&=s_{\{-n+1\}},&\omega(u_{\{-n+1\}},s_{-n})&=y_{\{-n+1\}},\\
\vdots&\vdots& \vdots&\vdots \\
\end{array}
\end{equation}
Therefore, a bi-infinite sequence   $\bm{y}=\{y_i\}_{i\in \Z}$ is said to be generated by the  WSHE $M=(U,S,Y,\nu,\omega)$
when there is an state $s_0$ and a sequence of inputs $\{u_i\}_{i\in \Z^+}$ such that $\{y_i\}_{i\in \Z^+}$
is obtained as equation (\ref{eq:d_system+}), and there are  $\{s_i\}_{i\in \Z^-}$, $\{u_i\}_{i\in\Z^-}\cup \{u_0\}$
such that $\{y_i\}_{i\in \Z^-}$ satisfies the equation (\ref{eq:d_system-}). 

Also it can be seen that the WSHE is linear and time invariant.

\begin{figure}
 \includegraphics[width=10cm]{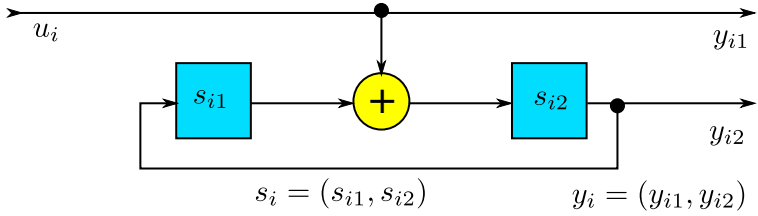}
\caption{The WSHE $M=(\Z_2,\Z_2^2,\Z_2^2,\nu,\omega)$ of the Example \ref{ex:binary_encoder}}
\label{fig:fsm_systematic}
\end{figure}

\begin{example}\label{ex:binary_encoder}
 Consider the encoder of the Figure \ref{fig:fsm_systematic}. This encoder is a WSHE.
 The inputs group is $U=\Z_2$, the states group is $S=\Z_2^2$
 and the output group is $Y=\Z_2^2$. The group extension is $U\boxtimes S=\Z_2\oplus \Z_2^2$.
 The next state homomorphism $\nu:\Z_2\oplus \Z_2^2\rightarrow \Z_2^2$ 
 is $\nu(u,s)=\nu(u,s_1,s_2)=(s_2,u+s_1)$ and
 the encoder homomorphism $\omega:\Z_2\oplus \Z_2^2\rightarrow \Z_2^2$ is $\omega(u,s)=\omega(u,s_1,s_2)=(u,s_2)$.
 This is a systematic convolutional which encodes sequences of single bits of $\Z_2=U$ in
 sequences of bit pairs of $\Z_2^2=Y$.
\end{example}
A finite sequence of bits $\{u_i\}_{i=1}^n$ is encoded initializing the encoder at state $s_0=(s_{01},s_{02})=00$. Then,
by using equation (\ref{eq:d_system+}) is obtained the encoded sequence $\{y_i\}_{i=1}^n$.
At the same moment $n$, when the last pair $y_n=(y_{n1},y_{n2})$ as $y_n=\omega(u_n,s_{n-1})$, the state
of the WSHE goes to $s_n=\nu(u_{n-1},s_{n-1})$. In practice, each time the encoding process is done, the encoder
state must be cleared, that is, it must be returned to state $00$. For that, it
may be necessary to add extra \text{padding} input bits $\{u_i\}_{n+1}^m$, $m>n$,
in such a way that  $00=\nu(u_m,\nu(u_{m-1},\dots,\nu(u_{n+1},s_n)))$. 
It is always possible to find, for this encoder
$M=(\Z_2,\Z_2^2,\Z_2^2,\nu,\omega)$, these extra padding bits, for any $\{u_i\}_{i=1}^n$.
For instance consider the input bits $\{u_i\}_{i=1}^7$ =
$\{0,1,1,1,0,1,0\}$, the sequence of states of the WSHE is
$\{s_i\}_{i=1}^7$ = $\{00,01,11,10,01,11,11\}$, whereas the encoded sequence is $\{y_i\}_{i=1}^7$ = $\{00,11,11,10,01,11,01\}$.
In this case the padding input extra bits are $\{u_8=1,u_9=1\}$ because $s_8=\nu(1,11)=10$ and $s_9=\nu(1,10)=00$. Hence,
$y_8=\omega(1,11)=10$, $y_9=\omega(1,10)=10$. Therefore $\{u_i\}_{i=1}^7$, padded as $\{u_i\}_{i=1}^9$, can be extended to the
bi-infinite sequence $\bm{u}=\{u_i'\}_{i\in \Z}$ where
$u_i'=\begin{cases} 0; i\leq 0\\  u_i;i\in\{1,2,\dots,9\}\\
0;i\geq 10 \end{cases}$ to produce the codeword $\bm{y}=\{y_i'\}_{i\in \Z}$ where 
$y_i'=\begin{cases} 00; i\leq 0\\  y_i;i\in\{1,2,\dots,9\}\\
00;i\geq 10 \end{cases}$, whereas that the state bi-infinite sequence is 
 $\{s_i'\}_{i\in \Z}$ such that $s_i'=\begin{cases} 00; i\leq -1\\  s_i;i\in\{0,1,2,\dots,9\}\\
00;i\geq 9 \end{cases}$ \lqqd

In order to generalize, for any WSHE, the input symbols \textit{padding} method, made in the Example \ref{ex:binary_encoder},
to extend $\{y_i\}_{i=1}^n$ to a codeword $\bm{y}=\{y_i\}_{i\in Z}$, 
we 
need to take into account that the return to the $s_0=e_S$, where $e_S$ is the neutral
element of $S$, demands the existence of a sequence
$\{y_i\}_{i=n+1}^m$, $m>n+1$ such that $e_S=\nu(u_m,\nu(u_{m-1},\dots,\nu(u_{n+1},s_{n+1})))$. 
A sufficient condition which guarantee the existence of padding input symbols that allowing the return
to zero state is that the group code $\mathcal{C}$ be controllable.

\section{Controllable group codes}
Each codeword of a group code satisfies the definition of a trajectory of a Dynamical System
in the sense of Willems \cite{willems}. From this each group code $\mathcal{C}$ is a dynamical system. In this context, the encoder
$M=(U,S,Y,\nu,\omega)$ is a \textit{realization} of $\mathcal{C}$, \cite{trott, loeliger, fagnani}.
Given a codeword $\bm{y}$  and a set of consecutive indices $\{i,i+1,\dots,j-1,j\}= [i,j]$, 
the projection of the codeword over these indices will be $\bm{y}\vert_{[i,j]}=\{y_i,y_{i+1},\dots,y_j\}$.
Analogously  $\bm{y}\vert_{[i,j)}=\{y_i,y_{i+1},\dots,y_{j-1}\}$,  $\bm{y}\vert_{[i,+\infty)}=
\{y_i,y_{i+1},\dots\}$ and so on. With this notation the \textit{concatenation}
of two codewords $\bm{y_1}$, $\bm{y_2} \in \mathcal{C}$ in the instant $j$ is  a sequence 
$\bm{y_1}\wedge_j\bm{y_2}$ defined by $ \begin{cases} 
(\bm{y_1}\wedge_j\bm{y_2})\vert_{(-\infty,j)}=\bm{y_1}\vert_{(-\infty,j)};\\
(\bm{y_1}\wedge_j\bm{y_2})\vert_{[j,+\infty)}= \bm{y_2}\vert_{[j,+\infty)}.
\end{cases}$ 

\begin{figure}
 \includegraphics[width=12cm]{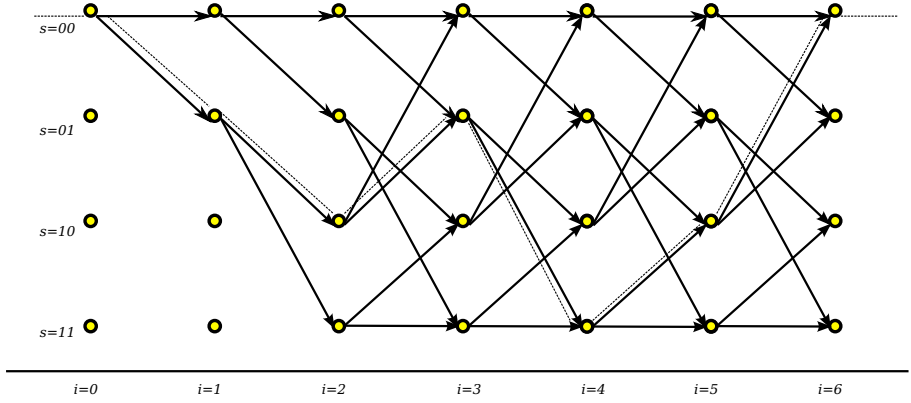}
\caption{Trellis diagram of the encoder $M=(\Z_2,\Z_2^2,\Z_2^2,\nu,\omega)$}
\label{fig:trellis}
\end{figure}

\begin{definition}\label{def:control}
If $L$ is an integer greater than one, then a group code $\mathcal{C}$ is said $L$-controllable if for  
any pair of codewords $\bm{y_1}$ and $\bm{y_2}$, there are a codeword $\bm{y_3}$ and one integer $k$
such that the concatenation $\bm{y_1}\wedge_k\bm{y_3}\wedge_{k+L}\bm{y_2}$ is a codeword of the 
group code $\mathcal{C}$. \cite{marcus,mittelholzer,willems}. \lqqd
\end{definition} 
It is said that a natural number $l>1$ is the index
of controllability of a group code $\mathcal{C}$ when $l=min\{L\;;\;\mathcal{C}\;\mbox{is}\; 
L-\mbox{controllable}\}$. Any applicable group code, for correction of errors of transmission and storage
of information, needs to have an index of controllability. Shortly, when a code has an index of controllability then
is said that it is controllable \cite{willems}. Clearly, a code $\mathcal{C}$ to be $L$-controllable is a sufficient condition for
$\mathcal{C}$ to be controllable. 

\subsection{Trellis of a group code produced by a WSHE}
The triplets $(s,\omega(u,s),\nu(u,s))$ of the set $\{\Psi(u,s)\}_{(u,s)\in U\boxtimes S}$, where
$\Psi$ is defined by (\ref{eq:Psi}), can be represented graphically. In the context of Graph Theory,
\cite{diestel}, they are called \textit{edges} whose vertexes set is  $S$ and the graph
is called \textit{state diagram} labeled by $\omega(u,s)$.
In the Figure \ref{fig:trellis} the full state diagram of the code generated by the FSM
$M=(\Z_2,\Z_2^2,\Z_2^3,\nu,\omega)$, from Example 2, is shown between the times 2 and 3 also it is 
repeated between the times 3 and 4 and so on until times 4 and 5.
In the context of Coding
Theory the elements of $\{\Psi(u,s)\}_{(u,s)\in U\boxtimes S}$ are called \textit{transitions} or \textit{branches}. 
The expansion in time of the state diagram is called \textit{trellis diagram}. This is made by concatenating at each time unit
separate state diagram. For two consecutive time units $i$ and $i+1$, the transitions $b_i=(s_i,\omega(u_{i+1},s_i),\nu(u_{i+1},s_i))$  and 
$b_{i+1}=(s_{i+1},\omega(u_{i+2},s_{i+1}),\nu(u_{i+2},s_{i+1}))$ are said concatenated when $s_{i+1}=\nu(u_{i+1},s_{i})$.
Hence a bi-infinite \textit{trellis path} of transitions is a sequence $\bm{b}=\{b_i\}_{i\in \Z}$ such that   $b_i$ and
$b_{i+1}$ are concatenated for each $i\in \Z$. The set of trellis paths form the trellis diagram. Since
each codeword $\bm{y}$ passes only by one state $s$ at each unit of time, then the relation between the codewords 
$\bm{y}$ and paths $\bm{b}$ is bijective.
Again from Example 2, consider the inputs sequence $\{u_i\}_{i\in \Z}$, such that $u_1=1,u_2=0,u_3=0,u_4=1,u_5=1,u_6=1$
and $u_i=0$ for all $i\in \Z-\{1,2,3,4,5,6\}$
The response  path $\bm{b}=\{b_i\}_{i\in \Z}$ is such that $b_0=(00,11,01)$, $b_1=(01,00,10)$
$b_2=(10,01,01)$, $b_3=(01,11,11)$, $b_4=(11,10,10)$, $b_5=(10,10,00)$
and $b_i=(00,000,00)$ for all $i\in \Z-\{0,1,2,3,4,5\}$. This response path is shown by a traced line in Figure \ref{fig:trellis}.

\begin{definition}\label{def:connection}
  Two states $s$ and   $r$ are said
 \textit{connected} when there are a path $\bm{b}$ and indices $i,j\in \Z$ such 
that $\bm{b}\vert_{[i,j]}=\{b_i,b_{i+1},\dots,b_j\}$
with $b_i=(s_i,\omega(u_{i+1},s_i),\nu(u_{i+1},s_i))$ and $b_j=(s_j,\omega(u_{j+1},s_j),\\ \nu(u_{j+1},s_j))$
such that $s=s_i$ and $r=\nu(u_{j+1},s_j)$. $\square$
\end{definition}

\begin{theorem}\label{theo:noncontrol}
Let $\mathcal{C}$ be a group code produced by the encoder $M=(U,S,Y,\nu,\omega)$.
If there are two states $s\in S$  and $r\in S$ for which there is not a finite path of transitions connecting them
then $\mathcal{C}$ is \textbf{non-controllable}.
\end{theorem}
\textbf{Proof.-}On contrary there is $l>1$ such that $l$ is the controllability index of $\mathcal{C}$.
Let $\bm{y_1}$ be one codeword passing by the state $s$ at time $k$, let $\bm{y_2}$ be
a codeword   passing by the state $r$ at time $k+L$, $L\geq l$. There must exist $\bm{y_3} \in \mathcal{C}$
with its respective path $\bm{b_3}$ such that $\bm{y_3}\vert_{(-\infty,k)}=\bm{y_1}\vert_{(-\infty,k)}$ 
and $\bm{y_3}\vert_{[k+L,+\infty)}=\bm{y_2}\vert_{[k+L,+\infty)}$  and $\bm{b_3}\vert_{(k,k+L]}$,
a finite path, connecting $s$ and $r$. Contradiction. \lqqd
 Equivalently, we can say that two states $s$ and $r$ are connected when there 
is a finite sequence of inputs
$\{u_i\}_{i=1}^n$ such that 
\begin{equation}\label{eq:connection}
r=\nu(u_n,\nu(u_{n1},\dots\nu(u_2,\nu(u_1,s))\dots)). 
\end{equation}

\section{The WSHE $(U,S,Y,\nu,\omega)$ with $U\boxtimes S$ abelian and $U=\Z_p$ }

In this section we will present the main results of this paper. We will show step by step that if we want
to construct WSHE, producing controllable codes, from an abelian group extension $\Z_p\boxtimes S$ 
 with $\Z_p$ being a $p$-prime cyclic group $\Z_p=\{1,2,\dots,p-1,p\}$,
then $S$ must be of the form $S=\Z_p^m$. This result is complementary with the one presented in
\cite{arpasi_mpe11} where it has been shown that the WSHE with  non-abelian extension  extension $\Z_p\boxtimes S$ 
produces non-controllable codes.

\begin{lemma}\label{lem:abel1}
 If $\Z_p\boxtimes S$ is abelian then $S$ is abelian.
\end{lemma}
\textbf{Proof.-} From the equation (\ref{eq:operation}) and Theorem \ref{theo:phi_nonabel}:
 $(u_1,s_1)*(u_2,s_2) = (u_1+u_2+\xi(s_1,s_2),s_1s_2)$. On the other hand, 
$(u_2,s_2)*(u_1,s_1) = (u_2+u_1+\xi(s_2,s_1),s_2s_1)$, hence $s_1s_2=s_2s_1$. Thus $S$ must be abelian. \lqqd

\begin{lemma}\label{lem:abel.cyclic}
 The abelian extension $\Z_p\boxtimes \Z_m$ either is isomorphic to the direct product $\Z_p\oplus \Z_m$ or
it is isomorphic to the cyclic group $\Z_{pm}$.
\end{lemma}
\textbf{Proof.-} Consider the element $(1,0)\in \Z_p\boxtimes \Z_m$. By the Theorem \ref{theo:phi_nonabel}, about the equation 
(\ref{eq:phi}), $(1,0)^2=(1+1+\xi(0,0),0)$. Now, by the equation (\ref{eq:xi}), $\xi(0,0)=0$. Thus, $(1,0)^2=(2,0)$ and in general
$(1,0)^n=(n,0)$, for any $n\in \{1,2,\dots,p-1\}$. Therefore  $H=\{(1,0),(2,0),\dots,(p-1,0),(0,0)\}$ is a cyclic
subgroup isomorphic with $\Z_p$.
On the other hand consider the element $(0,1)\in \Z_p\boxtimes \Z_m$. If $(0,1)^m=(0,0)$, then the subgroup 
$K=\{(0,1),(0,2),\dots,(0,m-1),(0,0)\}$ is isomorphic with $\Z_m$ and $H\cap K=\{(0,0)\}$. In this case, in accordance
with the Theorem 2.29, pg 40 of \cite{rotman} $\Z_p\boxtimes \Z_m$ must be isomorphic with the direct product $\Z_p\oplus \Z_m$.
In the case of $(0,1)^m= (u,0)$, with $u\neq 0$ we have that $u$ is a generator of $\Z_p$, then $((0,1)^m)^p= (u,0)^p=(0,0)$
with $((0,1)^m)^i\neq(0,0)$ for $0<i<p$. Therefore, $(0,1)^m=(u,0)$ implies that $\Z_p\boxtimes \Z_m$ is isomorphic with the cyclic group
$\Z_{pm}$ \lqqd

\begin{theorem} \label{theo:states_series}
Given the WSHE $(\Z_p,S,Y,\nu,\omega)$,  with $\Z_p\boxtimes S$ abelian, consider the family 
of state subsets $\{S_i^{(+)}\}$ and $S_0$, recursively defined by;
\begin{equation}\label{eq:states_series}
\begin{array}{lcl}
S_0&=&\{e_S\};\;e_S \mbox{ is the neutral element of } S\\
S_1^{(+)}&=&\{\nu(u,s)\;;\;u\in \Z_p,s\in S_0\}\\
S_2^{(+)}&=&\{\nu(u,s)\;;\;u\in \Z_p,s\in S_1^{(+)}\}\\
\vdots&\vdots&\vdots\\
S_i^{(+)}&=&\{\nu(u,s)\;;\;u\in \Z_p,s\in S_{i-1}^{(+)}\}, i\geq 1\\
\vdots&=&\vdots\\
\end{array}
\end{equation}
then;
\begin{enumerate}
\item Each $S_i^{(+)}$ is a normal subgroup of $S$.
\item If $S_{i-1}^{(+)}=S_i^{(+)}$ then  $S_i^{(+)}=S_{i+1}^{(+)}$.
\item If the group code is controllable then  $S=S_k^{(+)}$ for some $k<+\infty$.
\end{enumerate}
\end{theorem}

\textbf{Proof.-} 
\begin{enumerate}
\item By induction, consider $r,s\in S_i^{(+)}$, Since $\nu$ is surjective, there exist $(u_1,s_1)$ and $(u_2,s_2)$
with $s_1,s_2\in S_{i-1}^{(+)}$ and $u_1,u_2\in \Z_p$
such that $r=\nu(u_1,s_1)$ and $s=\nu(u_2,s_2)$. Hence, $sr=\nu(u_3,s_1s_2)$, $u_3\in \Z_p$ and thus
$sr\in S_i^{(+)}$.
\item Given $s\in S_{i+1}^{(+)}$ there are $r\in S_i^{(+)}$ and $u\in \Z_p$ such that $\nu(u,r)=s$. Since $S_i^{(+)}=S_{i-1}^{(+)}$, $r\in S_{i-1}^{(+)}$.
Hence $\nu(u,r)=s\in S_i^{(+)}$.
\item On the contrary, there is $s\in S$ such that $s\not\in S_k^{(+)}$, for any $k\in \mathbb{N}$. Then, 
the neutral state $e_S\in S_k^{(+)}\subset S$
and $s$ are not connected by any finite trellis path. By the Theorem \ref{theo:noncontrol}, the group code would be
 non-controllable. 
\end{enumerate}
\lqqd
In the Figure \ref{fig:trellis}, $S_0=\{00\}$, $S_1^{(+)}=\{00,10\}$, $S_2^{(+)}=\{00,10,01,11\}=S$, hence the code is controllable.

\begin{lemma}\label{lem:s_past0}
Let $S_1^{(-)}$ be the full one-time past of the neutral 
state $s_0=e\in S$, precisely defined by 
\begin{equation}\label{eq:s_past}
 S_1^{(-)}=\{s\in S\;;\;\nu(u,s)=e_S\,,\mbox{ for some } u\in \Z_p\}.
\end{equation}
Then $S_1^{(-)}$ is a normal subgroup of $S$ and $|S_1^{(-)}|=|S_1^{(+)}|=p$
\end{lemma}

\textbf{Proof.-}
Consider the kernel of $\nu$ and the kernel of the second projection  $\pi_2(u,s)=s$. Both $\nu$ and $\pi_2$ are
surjective homomorphisms, then, by  the fundamental homomorphism Theorem, $\frac{\Z_p\boxtimes S}{ker(\nu)}\cong S$ and
 $\frac{\Z_p\boxtimes S}{ker(\pi_2)}\cong S$. Hence, $|ker(\nu)|=|ker(\pi_2)|$.
Now, $ker(\pi_2)=\Z_p\boxtimes \{e_S\}$ yields $|ker(\pi_2)|=p$. If $|ker(\pi_2)|=1$, we would have the trivial case $|S|=1$.
Therefore, the statement of the Lemma is satisfied
noticing that $S_1^{(-)}=ker(\nu)$ and $S_1^{(+)}=ker(\pi_2)$.
\lqqd

\begin{lemma}\label{lem:s_past}Let $M=(\Z_p,S,Y,\omega,\nu)$ be a WSHE with $p$ prime. 
Let $\{S_i^{(+)}\}_{i\geq 1}$ be the sequence defined by equation (\ref{eq:states_series}), and let  
$S_1^{(-)}$ be the subgroup defined by equation (\ref{eq:s_past}), then:
\begin{enumerate}
 \item If there are $s\neq e_S$, with $s\in S_1^{(-)}\cap S_i^{(+)}$, then $S_1^{(-)}\subset S_i^{(+)}$.
\item If $S_1^{(-)}\subset S_i^{(+)}$ then $\nu(\Z_p,S_1^{(-)})\subset S_i^{(+)}$.
\end{enumerate}
\end{lemma}
\textbf{Proof.- }
\begin{enumerate}
 \item By Lemma \ref{lem:s_past0}, $|S_1^{(-)}|=|S_1^{(+)}|=p$. Then $S_1^{(-)}=\{s,s^2,\dots,s^{p-1},s^p=e_S\}\subset S_1^{(-)}\cap S_i^{(+)}$.
\item Given $r\neq e_S$ such that $r\in S_i^{(+)}\cap S_1^{(-)}$ suppose there is some $u\in \Z_p$ such that $\nu(u,r)=s\not\in S_{i}^{(+)}$. 
For the subgroup $S_1^{(+)}=\{s_0,s_1=\nu(u_1,e_S),s_2=\nu(u_2,e_S),\dots,s_{p-1}=\nu(u_{p-1},e_S)\}$, we have that $sS_1^{(+)}$ is a coset where 
each element is $\nu(u,r)\nu(u_i,e_S)=\nu(u^\prime,r)$, for some $u^\prime\in \Z_p$. Hence $sS_1^{(+)}=\{\nu(\Z_p,r)\}$ with $sS_1^{(+)}\cap S_i^{(+)}=\emptyset$. 
But, since $r\in S_1^{(-)}$ there is at least one $u_0\in \Z_p$ such that $\nu(u_0,r)=e_S$, in contradiction with $sS_1^{(+)}\cap S_i^{(+)}=\emptyset$. \lqqd
\end{enumerate}

\begin{definition}\label{def:index}
 Given a finite group $G$ and a subgroup $H\subset G$, the \textbf{index} of $H$ in $G$, denoted
by $[G:H]$ is the number of different cosets of $H$ in $G$ and $[G:H]=\frac{|G|}{|H|}$, \cite{rotman}. \lqqd
\end{definition}

\begin{theorem}\label{theo:si_mustbe_p} Let $M=(\Z_p,S,Y,\omega,\nu)$ be a WSHE with $p$ prime,
 then each $S_i^{(+)}$ of (\ref{eq:states_series}) must be a $p$-group. 
\end{theorem}
\textbf{Proof.- }
By induction over $i$. For $i=1$, by Lemma \ref{lem:s_past0},  $[S_1^{(+)}:S_0]=p$.
Now suppose there is a natural number $k>1$ such that $[S_i^{(+)}:S_{i-1}^{(+)}]=p$, for all $i\leq k$. Then,
the subgroup $S_k^{(+)}$ has $p^k$ elements, each one with order $p^i$, $i\leq k$.
If $p>[S_{k+1}^{(+)}:S_k^{(+)}]>1$ then $[S_{k+1}^{(+)}:S_k^{(+)}]=m=q_1^{r_1}q_2^{r_2}\dots q_t^{r_t}$, where each $q_i$ is a prime
and $q_j< p$. There must be an element $s\in (S_{k+1}^{(+)}-S_k^{(+)})$, the difference set, such that $s^{q_1}=e_S$.
Let $u\in \Z_p$ and $r\in S_k^{(+)}$ be such that $\nu(u,r)=s$, then 
$(\nu(u,r))^{q_1}=\nu(u_1,r^{q_1})=s^{q_1}=e_S$. Hence $r^{q_1}\in S_1^{(-)}\cap S_k^{(+)}$.\\
If $r\neq e_S$ then $r^{q_1}\neq e$, because $q_1< p$. By Lemma \ref{lem:s_past}, $S_1^{(-)}\subset S_k^{(+)}$
and $\nu(u,r)=s\in S_k^{(+)}$, a contradiction.\\
If $r=e_S$ then $\nu(u,r)=s\in S_1^{(+)}\subset S_k^{(+)}$, also a contradiction. \lqqd
\begin{corollary}
 If $[S_k^{(+)}:S_{k-1}^{(+)}]=p$ then $\nu(u,s)\in (S_k^{(+)}-S_{k-1}^{(+)})$, the difference set, for all $s\neq e$.
\end{corollary}

\begin{corollary}
 If the code is controllable then $|S_i^{(+)}|=p^i$.
\end{corollary}

\begin{lemma}\label{lem:s_past1}
 If $S_1^{(-)}\cap S_i^{(+)} \neq \{e\}$ for some $S_i^{(+)}\neq S$, then the code produced by the WSHE $M=(\Z_p,S,Y,\omega,\nu)$ is non-controllable.
\end{lemma}
In accordance with item 1 of the  Lemma \ref{lem:s_past}, $S_1^{(-)}$ is a subset of $S_i^{(+)}$. By Theorem \ref{theo:si_mustbe_p}, 
 $S_i^{(+)}$ is a $p$-group which have $S_1^{(+)}$ and  $S_1^{(-)}$ as subgroups of order $p$. 
Since any $p$-group has only one subgroup with order $p$, then $S_1^{(+)}=S_1^{(-)}$. Again, 
by the item 2 of the  Lemma \ref{lem:s_past}, $\nu(\Z_p,S_1^{(+)})\subset S_1^{(+)}$. Therefore, considering any $s\in S$ such that
$s\not\in S_1^{(+)}$ we have that there is not any finite path connecting the neutral element $e\in S_1^{(+)}\subset S$ and $s$. In accordance 
with the Theorem \ref{theo:noncontrol}, the code is non-controllable
 \lqqd

\begin{theorem}\label{theo:cyclic}
Consider the WSHE $M=(\Z_p,S,Y,\nu,\omega)$ defined over the abelian extension $\Z_p\boxtimes S$. 
If $i\geq 2$ and $S_i^{(+)}$ is cyclic then $S_{i+1}^{(+)}$ is cyclic.
\end{theorem}
\textbf{Proof.-} Since any subgroup of a cyclic group is also cyclic, then $S_{i-1}^{(+)}$ must be cyclic and isomorphic with 
$\Z_{p^{i-1}}$ (Corollary of Theorem \ref{theo:si_mustbe_p}). Then, either $S_i^{(+)}\cong \Z_{p^{i-1}}$ or $S_i^{(+)}\cong \Z_{p^i}$.
If $S_i^{(+)}\cong \Z_{p^{i-1}}$, then $S_j\cong \Z_{p^{i-1}}$ for all $j\geq i$. By
Theorem \ref{theo:states_series}, the WSHE $M=(\Z_p,S,Y,\nu,\omega)$ would produce a non-controllable code. 
Thus, $S_i^{(+)}$ must be isomorphic to $\Z_{p^i}$.\\
Now, suppose $S_{i+1}^{(+)}=\nu(\Z_p\boxtimes S_i^{(+)})$ is not cyclic, then by Lemma \ref{lem:abel.cyclic}, 
$S_{i+1}^{(+)} \cong \Z_p\oplus \Z_{p^i}$.
For the sake of clarity let us write $S_{i+1}^{(+)}$ as $S_{i+1}^{(+)}=\Z_p\oplus \Z_{p^i}$. Then each element of $S_{i+1}^{(+)}$ is a pair
$(x,y)$ with $x\in \Z_p=\{0,1,\dots,p-1\}$ and $y\in \Z_{p^i}=\{0,1,\dots,p,\dots,2p,\dots,p^i-1\}$.
Consider the pair $(0,p)\in S_{i+1}^{(+)}$. The order of $(0,p)$ is $p^{i-1}$, therefore $S_{i-1}^{(+)}$ is generated by 
$(0,p)$. Now, choose any $(k_1,k_2)\in (S_i^{(+)}-S_{i-1}^{(+)})$ and let $(x,y)$ be such that  $(x,y)=\nu(0,(k_1,k_2))$. On one side we have
that $(x,y)^p=(0,py)\in S_{i-1}^{(+)}$. On the other side $(x,y)^p=(\nu(0,(k_1,k_2)))^p=\nu((0,(k_1,k_2))^p)=\nu(u,(0,k_2p))$.
But by the Theorem \ref{theo:si_mustbe_p}, $\nu(u,(0,k_2p))$ must be in $(S_i^{(+)}-S_{i-1}^{(+)})$, a contradiction.  \lqqd 
\begin{theorem}
 Consider the WSHE $M=(\Z_p,S,Y,\nu,\omega)$ defined over the abelian extension $\Z_p\boxtimes S$. If $S$ is cyclic then
the code is not controllable
\end{theorem}
\textbf{Proof.-}If $S$ is cyclic then there is a unique subgroup of $\Z_p\boxtimes S$, with order $p$. This means $S_1^{(-)}=S_1^{(+)}$. 
By Lemma \ref{lem:s_past}, 
$S_i^{(+)}\subset S_1^{(+)}$ for all $i$. By Lemma \ref{lem:s_past1} the code is non controllable. \lqqd
\begin{theorem}\label{theo:main}
If the code produced by the WSHE $M=(\Z_p,S,Y,\nu,\omega)$ defined over the abelian extension $\Z_p\boxtimes S$ 
is controllable, then $S$ must be isomorphic with $\Z_p^j$ for some natural $j\geq 1$
\end{theorem}
\textbf{Proof.-} If the code is controllable then $S_1^{(+)}\cong \Z_p$. Now, by Theorem \ref{theo:cyclic}, if
the code is controllable, then
for $i\geq 2$, $S_{i}^{(+)}=\nu( \Z_p\boxtimes S_{i-1}^{(+)})\cong\Z_p\oplus \Z_{p^{i-1}}$. \lqqd
\section{Conclusions}
The main result of this article, which is the Theorem \ref{theo:main}, was shown by using properties
of the states subgroups $S_i^{(+)}$  and $S_1^{(-)}$ defined by equations  (\ref{eq:states_series})
and (\ref{eq:s_past}) respectively. Immediate related problem is the study on control conditions
for the case in which the WSHE is defined over abelian extensions of the form $\Z_p^n\boxtimes S$ or
$\Z_{pn}\boxtimes S$, where $\Z_p^n=\overbrace{\Z_p\oplus\Z_p\oplus\dots\oplus\Z_p}^{n-times}$
and $\Z_{pn}$ is the cyclic group of order $pn$, $p$ prime, $n\geq 1$ is a natural number. 
Since the Lemma \ref{lem:abel1}, it is clear that $S$ must be abelian. Then, how would be the structure
of $S$?, how would be the structures of the sets $S_i^{(+)}$  and $S_1^{(-)}$?,
are questions that, we think, must be answered in order to get some control conditions for the group codes 
produced by a WSHE defined over extensions of the form $\Z_p^n\boxtimes S$ or
$\Z_{pn}\boxtimes S$.

\bibliography{../biblioteca}
\bibliographystyle{unsrt}

\end{document}